- Manuscript -

**A pretest-posttest pilot study for augmented reality-based physical-cognitive training in community-dwelling older adults at risk of mild cognitive impairment**

Sirinun Chaipunko[a], Watthanaree Ammawat[a*], Keerathi Oanmun[b], Wanvipha Hongnaphadol[c], Supatida Sorasak[a], Pattrawadee Makmee[d]

[a]Division of Occupational Therapy, Faculty of Physical Therapy, Mahidol University,

[b]Physical Therapy Center, Faculty of Physical Therapy, Mahidol University

[c]Faculty of Management Sciences, Kasetsart University

[d]Department of Research and Applied Psychology, Faculty of Education, Burapha University

*Corresponding author. 999 Phuttamonthon 4 Road, Salaya, Nakhon Pathom 73170 Thailand

E-mail address: watthanaree.ama@mahidol.ac.th

*Abstract*

As cognitive interventions for older adults evolve, modern technologies are increasingly integrated into their development. This study investigates the efficacy of augmented reality (AR)-based physical-cognitive training using an interactive game with Kinect motion sensor technology on older individuals at risk of mild cognitive impairment. Utilizing a pretest-posttest experimental design, twenty participants (mean age 66.8 ± 4.6 years, age range 60-78 years) underwent eighteen individual training sessions, lasting 45 to 60 minutes each, conducted three times a week over a span of 1.5 months. The training modules from five activities, encompassing episodic and working memory, attention and inhibition, cognitive flexibility, and speed processing, were integrated with physical movement and culturally relevant Thai-context activities. Results revealed significant improvements in inhibition, cognitive flexibility, accuracy, and reaction time, with working memory demonstrating enhancements in accuracy albeit not in reaction time. These findings underscore the potential of AR interventions to bolster basic executive enhancement among community-dwelling older adults at risk of cognitive decline.

*Introduction*

Ageing is characterized by progressive physiological deterioration, which is primary risk factor of changing in the declining of nervous system with consequent alterations both physical and mental functions [1, 2]. The critical hallmarks of neurodegenerative symptom including genomic instability and increase of toxic-protein aggregation, which have been associated with various neurodegenerative disease such as Alzheimer's disease (AD), Parkinson's disease (PD) [1]. However, those aged 65 and older who actively engage in intellectual or cognitive activities such as reading book, playing board game, Mahjong or card game had reduced risk of dementia [3].

The underlying factors which delayed the declining of cognitive function in elderly are the individual variations of executive functions (EFs). These are the higher level of cognitive function especially in goal-directed behavior and oriented future behavior. Mainly, EFs were examined concerning inhibitory control of behavior, shifting or mental flexibility and working memory in updating novel information. These functions are commonly observed in everyday life situations [4]. Extensive numbers of studies were reported in the systemically analysis study that the significant benefits of physical activities on EFs in areas of working memory, mental flexibility, as well as in the activity of daily living (ADLs) performance in AD patients [5]. Additionally, physical activity is regarded as a vital factor influencing quality of life, with beneficial effects extending beyond physical health to encompass psychological well-being [6].

As Thailand's population ages, the prevalence of daily functional disabilities and cognitive decline among Thai older adults has been steadily increasing [7]. The maintenance of a healthy living environment for individuals involves integrating aspects such as lifestyle choices, dietary habits, attire, and the pursuit of happiness across generations [8]. The way of living in Thai population is bounded with native culture, which is the representation of national wisdom enclosing the intensive cultural atmosphere, a high historical artistic and scientific values [9]. However, the mostly wealthier families in rural areas of Thailand tend to be more restful and spend most of time to watching TV and enjoying social media, which is poorly sedentary behavior [10]. As evidenced, there exists a positive correlation between sedentary behavior and cognitive impairment among populations with conditions such as mild cognitive impairment or dementia [11]. However, older adults who engage in regular physical activity and maintain an active lifestyle in their daily routines can postpone declines in health and sustain functional performance for extended periods compared to those leading sedentary lifestyles [10].

Augmented reality (AR) and virtual reality (VR) technologies have been developed to facilitate direct interaction between users and their real-world environment. AR holds significant potential for applications in various everyday life situations, including real-world orientation (navigation), education, healthcare, entertainment, and social interaction [12]. Especially in cognitive aspect, the VR shows the potential effect to improve the cognitive rehabilitation for individuals with dementia [13] and older adults with mild cognitive impairment [14]. In addition, the cognitive-motor combined with technology-based intervention known as technology-based exergame interventions had a positive effect on cognitive function, inhibition, processing speed of motor response [15].

Accordingly, the integration of new technologies and media has become prevalent in the lifestyles of individuals, including older adults residing in rural areas of Thailand. This study aims to investigate the effectiveness of "Multiple Cognitive and Physical Activity Augmented Reality Training (MU-COPART)" as a home-based physical activity to enhance EFs for community-dwelling elderly people with mild cognitive impairment in Bangkok, Nakhon Pathom and Chonburi provinces of Thailand.

*Materials and Methods*

**Participants**

The community-dwelling older adults with mild cognitive impairment were included in this study with aged over 60 years old. All participants were the volunteer and member of the Subdistrict Health Promotion Hospital and senior citizen club in Bangkok, Nakhon Pathom and Chonburi provinces of Thailand. Participants meeting the following criteria were included: all individuals aged over 60 years exhibiting mild cognitive impairment as indicated by scores lower than 25 on the Montreal Cognitive Assessment (MoCA)-Thai-version 01 or lower than 20 on Behavioral Assessment of the Dysexecutive Syndrome (BADS). Additionally, participants were required to have normal vital signs, muscle strength in hand grip, proficiency in the Thai language, visual acuity scores exceeding 1 on the Freiburg Visual Acuity and Contrast Test (FrACT) [16] and normal color vision as assessed by a color blindness test. Participants with dementia or other neurological conditions, visual or hearing impairment affecting communication skills as well as those at risk of falling as determined by the timed up and go test, were excluded. The research ethics protocol was reviewed by the University Institutional Review Board, with approval granted under COA No. MU-CIRB 2021/191.2709. Prior to participating in the research project, all participants provided informed consent. The clinical trial was reviewed and approved by the TCTR Committee. The TCTR identification number is TCTR20220629004

(https://www.thaiclinicaltrials.org/show/TCTR20220629004).

**Development of the augmented reality technology program**

Multiple Cognitive and Physical Activity Augmented Reality Training, or so-called MU-COPART, an augmented reality technology program, was developed by a team of researchers to enhance the EFs of older adults in the community. The program focuses on home-based physical activity. The integration of Thai culture and practices into the local context was achieved through a developmental program created by Unity. This program consists of five activities: Sukjai Trekking, Fun Festival, Delicious Menu, Sea Pleasure, and Sequential Order. MU-COPART was developed as a home-based physical activity intervention aimed at enhancing EFs among community-dwelling older adults. The implementation of MU-COPART was facilitated using a laptop computer connected to a Kinect® camera. Prior to commencing the program, participants were presented with a choice of six avatars, as illustrated in Figures 1 (A) and 1 (B). The theoretical framework of the MU-COPART underwent thorough evaluation by a panel of five expert committees to ensure its validity. Content validity, assessed by the Content Validity Index (CVI), yielded a score of 0.96, indicating high quality in the operational procedures and physical movement activities aimed at enhancing EFs such as attentional control or inhibition, working memory, and mental flexibility within the Thai context.

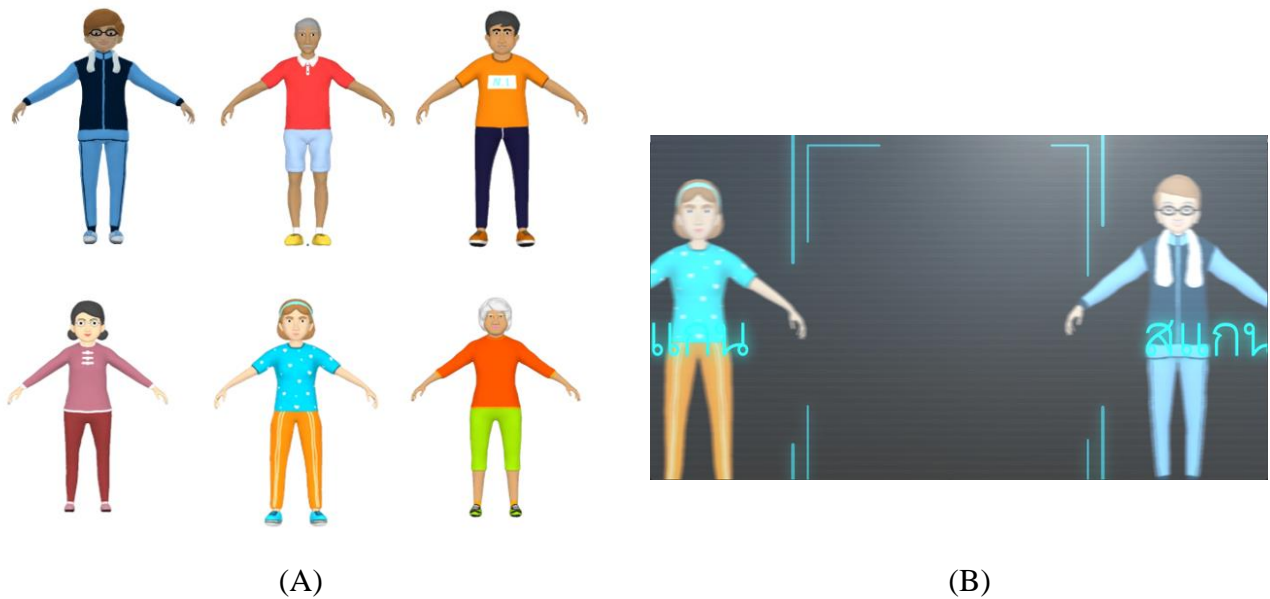

(A)                                                                                      (B)

Figure 1. (A) Six avatar characters, comprising three men and three women, were presented instead of an older person. (B) The setup screen for connecting the Kinect camera

**Study design**

A pre-posttest single cross-sectional study was conducted for this preliminary pilot study to investigate the effectiveness of MU-COPART implementation on EFs of older adult participants with mild cognitive impairment (MCI). MU-COPART was implemented as a home-based physical activity aimed at enhancing EFs in the community-dwelling older adults with MCI. The implementation consisted of 18 sessions over 6 weeks (3 sessions a week, each lasting 45-60 minutes) designed to enhance the individual EFs. The study design is illustrated in Figure 2.

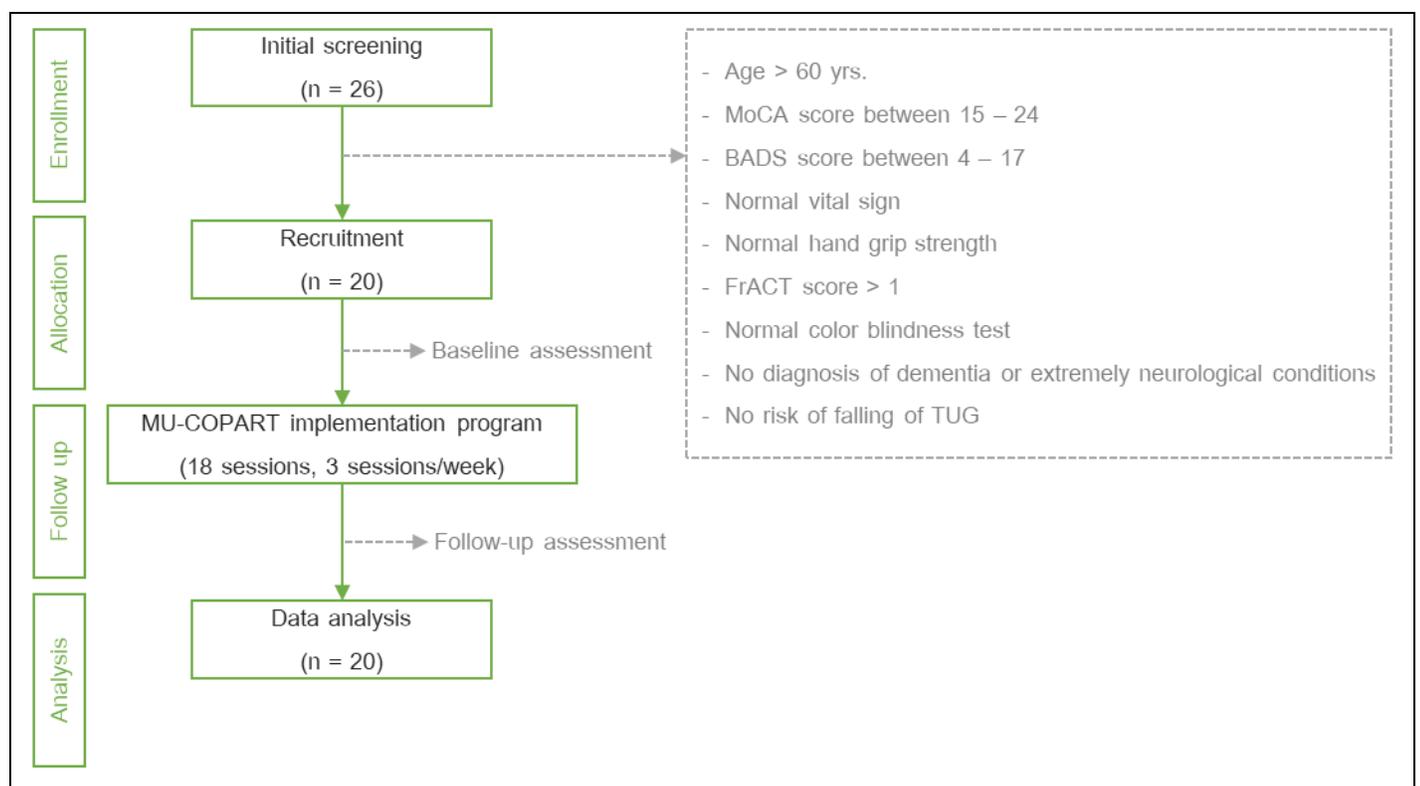

**Figure 2 Study chart flow**

**Baseline measurement of data collection**

*General data collection questionnaires*

The questionnaires were used to collect demographic information from individual participants. This included participant coding, residential province, year of birth, gender, age, date of birth, educational level, annual income, occupation, medical history, and individual physical activity habits.

*The Lawton Instrumental Activities of Daily Living Scale (L-IADLs)*

This assessment tool measures functional abilities related to instrumental-activities of daily living (IADLs) encompassing eight complex tasks necessary for community living [17]. These tasks include using the telephone, shopping, food preparation, housekeeping, laundry, using transportation, handling medications, and managing finances. The administration time for L-IADLs is 10- 15 minutes, enabling the identification of declines in physical and cognitive function in older people, even in those who may outwardly appear healthy.

*Timed Up and Go Test (TUG)*

This test was chosen to test the walking performance and risk of falling of participants. Each participant was instructed to walk forward and backward along a straight line for a distance of 3 meters and then return to a seated position. The time taken to complete the task, from the command 'Go' to sitting back down, was recorded [18].

*Grip strength test*

Jamar hand-grip dynamometers were used to measure the grip strength of all participants. Each participant was instructed to grip the power unit of the dynamometer twice, and the maximal force exerted by each participant was recorded. For individuals aged between 60 and 69 years, the normal hand grip strength typically falls within the range of 14.7 to 31.9 kilograms.

*Freiburg Visual Acuity and Contrast Test (FrACT)*

The FrACT is computer software used to assess the visual acuity of older people [19].

*Color blindness test*

The Ishihara plates was selected to screen the color blindness conditions. The plate including 38 picture pages. Each plate consisted of randomly arranged dots with distinct hidden numbers. The participants were asked to look on each plate and recognized the distinct hidden numbers. [20]

*Montreal Cognitive Assessment (MoCA) – Thai version 7.1*

The MoCA – Thai version 01 is developed for screening cognitive functional performance in Thai language. Each assessment session lasted for 10 minutes. The assessment evaluates six cognitive domains, with a maximum score of 30 points. These domains include EF, visuospatial, short-term memory, language, attention, concentration and working memory, and temporal and spatial orientation. A MoCA score between 17 and 24 indicates MCI, while a total score below 16 suggests dementia of Alzheimer's type (DAT)[21].

*Behavioral Assessment of the Dysexecutive Syndrome (BADS)*

The Behavioral Assessment of the Dysexecutive Syndrome (BADS) is employed to predict performance difficulties arising from dysexecutive syndrome in daily life activities. The maximum score on the BADS is 24 points, with overall classification spanning 7 levels of EF performance: impaired, borderline, low average, average, high average, superior, and very superior [22].

The BADS comprises 6 subtests and an additional questionnaire, as detailed below.

*Rule shift cards test;*

This test involves the use of 21 spiral-bound non-picture playing cards. Participants were tasked with correctly responding to a rule, shifting from the old rule to a novel one, and keeping track of the rule of the previous card and the current one. The time taken and the number of errors were recorded to determine a maximum score of 4 points.

*Action program test;*

This test was designed to present participants with a novel and practical task. Participants were assigned to develop their own plan of action to solve the problem at hand. To successfully tackle the problem, participants needed to devise a plan consisting of 5 sequential steps aimed at making the cork float to the top of the tall tube. The maximum score for the action program test is 4 points.

*Key search test;*

The participants were tasked with locating a lost key within a 100 squared-mm field on an A4 paper. Raw scores were determined based on four criteria: i) the participants' entry and exit from the field, ii) their searching pattern, iii) a single line search for the key, and iv) the continuation of the search line. The total raw score could reach 16 points, which was then converted to a 4-point profile score.

*Temporal judgement test;*

The participants were presented with four general questions related to daily life activities. Each question required a brief response regarding the time spent. Participants were asked to estimate the duration of four different activities, ranging from a few seconds to several years. The score for the temporal judgement test would be 4 points on the profile score.

*Zoo map test;*

The zoo map test evaluates participants' ability to plan their route to visit a series of designated locations on a zoo map. Participants attempted two trials, each with the same objective. The total raw score of the zoo map test would be 16 points, which was then converted into 4 points on the profile score.

*Modified six element test;*

The participants were instructed to complete three tasks: dictation, arithmetic, and picture naming. Each task was divided into two subtasks: A and B. Participants were required to complete all six subtasks. The total raw score for this section would be 6 points, which was then converted into 4 points on the profile score.

**MU-COPART implementation**

The MU-COPART is an augmented reality technology program specifically designed to enhance EFs in community-dwelling older adults. It is implemented as a home-based physical activity program. The physical tasks included in the MU-COPART program consist of walking, marching, arm swing, square-stepping, and lateral walking. Participants engaged in these five physical activities over 18 sessions spanning 6 weeks (3 sessions a week), with each session lasting between 45 and 60 minutes (see Figure 3). Each physical activity was designed with 3 different levels with 10 sessions in each level, except for the Delicious Menu activity, which has 5 sessions in each level. Each physical activity requires various performance tasks as described below.

*Sukjai Trekking*

The Sukjai Trekking was designed with consideration of environmental perception, featuring a greenery background aimed at facilitating individual inhibition and self-control skills. Participants were prompted to select a target picture corresponding to the audio vocabulary accompanied with a stimulus image (featuring basic animals or fruits). Completing each level of the activity requires arm swinging and walking.

*Delicious Menu*

The Delicious Menu was developed to feature familiar Thai traditional foods, akin to those found in daily home-cooked meals, such as crispy pork spicy salad, salty fried chicken, stir-fried morning glory, pad-thai, and tom yum kung. Engaging in this activity allows participants to enhance their working memory and recognition performance. The primary task required to complete this physical activity is the arm swinging.

*Fun Festival*

The Fun Festival was designed to improve mental flexibility and inhibition skills. Participants engage in marching to complete this physical activity. Inspired by familiar Thai cultural contexts such as Songkran, Chinese Lunar New Year, temple fairs, and Loy Krathong festivals, participants select pictorial stimuli that correspond to festival scenarios.

*Sea Pleasure*

The Sea Pleasure activity immerses participants in a seaside setting, featuring sea animals and marine activity equipment. Participants engage in square stepping while memorizing a target picture to complete this physical activity. This task aims to enhance participants' working memory and recognition performance.

*Sequential Order*

The sequential order was established based on numerical perception and sequencing to execute goal-directed movements, including marching and lateral walk. This physical activity aims to enhance the inhibition and self-control abilities of the individual.

Examples of using the MU-COPART and monitor screen are shown in Figures 3 and 4.

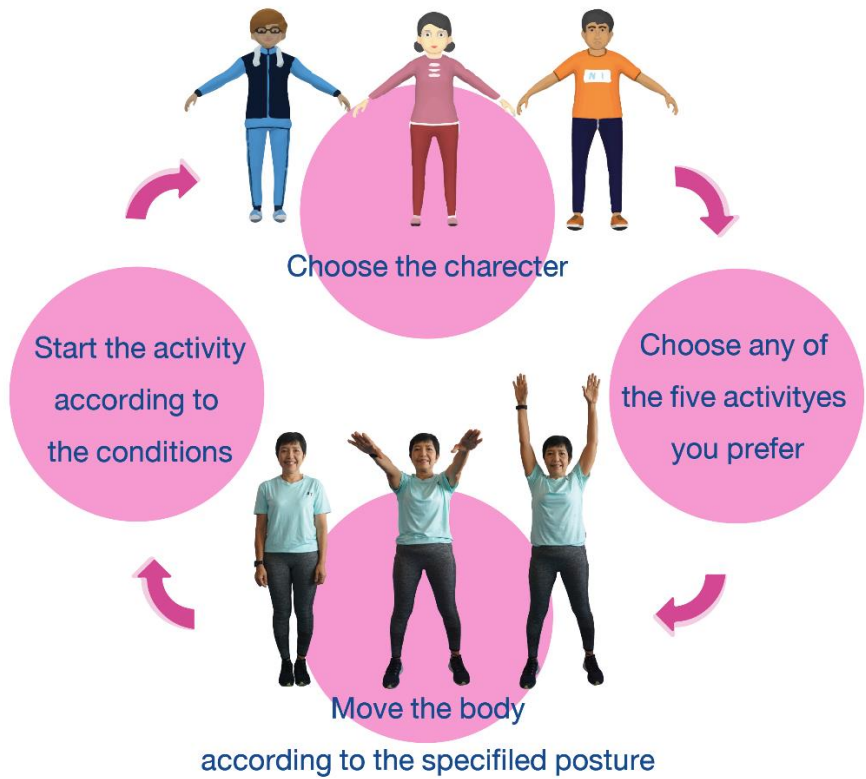

Figure 3 The sequence involved in utilizing the MU-COPART program and engaging in physical activity.

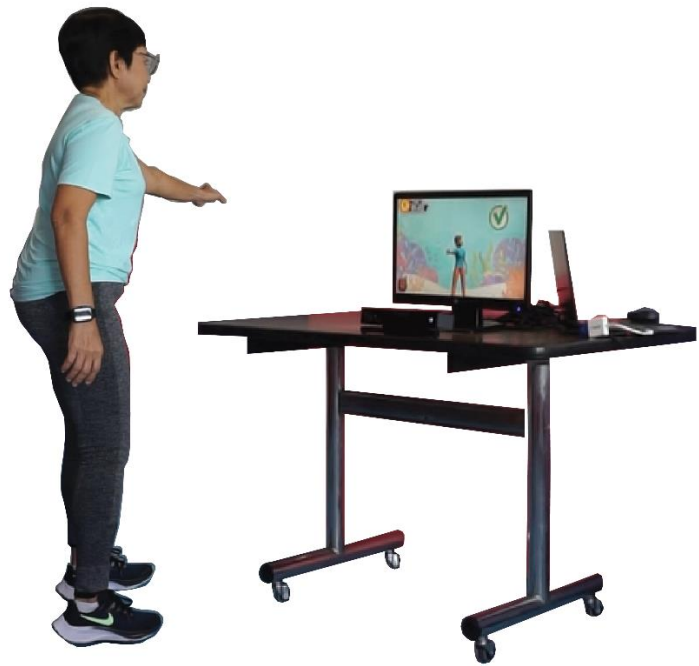

Figure 4 Engaging with the MU-COPART system.

**Instruments for measuring dependent variables.**

*The computer-based of advanced executive function evaluations*

The Psychology Experiment Building Language (PEBL) program, which is freely available as open-source software, is used as computer-based evaluations. The study incorporated three distinct tests (Flanker task, digit span test, Berg card sorting test-64), each aimed at evaluating EFs, described as follows.

*Flanker task*

The Flanker task, also known as the Eriksen Flanker task, is a method used to assess attention, specifically selective attention, and inhibition. Participants are instructed to press the left or right arrow key based on the direction of the target stimulus. [23]

*Digit span test*

The digit span test assesses verbal working memory. Participants are required to memorize a series of digits, which gradually increases in length, both in forward spans. Typically, individuals with normal working memory can memorize approximately 7 digits in forward order.

*Berg card sorting test-64*

This test was devised to assess mental flexibility, measuring individuals' capacity to solve dynamic problems with fixed rules. The test involves presenting four reference cards arranged in a single row on the monitor, along with a single stimulation card in a designated location on the computer monitor. Participants are then prompted to select a reference card that matches the stimulus card based on a series of predefined conditions.

**Data analysis**

Descriptive information, including the participants' residential province, gender, cognitive performance (assessed using MoCA and BADS-EF) was collected to assess the feasibility of attending all 18 sessions of the MU-COPART program. Changes in inhibition, working memory, and mental flexibility before and after participating in the MU-COPART program were compared using paired sample t-tests. Statistical analyses were conducted using IBM SPSS® Statistics software version 18 with a significance level set at 0.05.

*Results*

This pilot study aimed to investigate the effectiveness of the "Multiple Cognitive and Physical Activity Augmented Reality Training" (MU-COPART) influence on the EFs of older adult participants with MCI.

*Participant characteristics*

The demographic characteristics of the consenting participants are presented in Table 2. All 20 consenting participants volunteered and were members of the Subdistrict Health Promotion Hospital and senior citizen club in Bangkok, Nakhon Pathom and Chonburi provinces of Thailand. Most of the participants were female (n=14, 70%), with ages ranging from 60 to 78 years, and an average age of 66.8 years (SD=4.6). Cognitive performance, as measured by MoCA score, ranged from 15 to 24, with an average score of 19.95 (SD=2.74), while BADS-EF scores ranged from 4 to 17, with an average score of 12.35 (SD=3.28).

**Table 2 Baseline characteristics of the participants (n=20)**

| Characteristics | Mean (SD) |
|---|---|
| Gender (n/percentage) | |
| - Male | 6 / 30% |
| - Female | 14 / 70% |
| Age range: 60 – 78 yrs. | 66.8 (4.6) |
| Montreal Cognitive Assessment (MoCA) ranged from 15 to 24 | 19.95 (2.74) |
| Behavioral Assessment of the Dysexecutive Syndrome (BADS) ranged from 4 to 17 | 12.35 (3.28) |

The effectiveness of MU-COPART implementation following participants' engagement in 5 physical activities for 18 sessions over 6 weeks (3 sessions a week, lasting 45-60 minutes each) is presented in Table 3. The table displays the mean, standard deviation (SD), minimum, maximum, skewness, and kurtosis values of EF parameters of the participants. Skewness values ranged from -1.39 to 4.07, indicating a normal distribution, while kurtosis values were mostly within the normal range, except for the kurtosis value of response time in the digit span test post MU-COPART program. However, the majority of skewness and kurtosis values were indicative of normal distribution, falling within the defined range of skewness between (-2 to +2) and kurtosis (-7 to +7) for normal data [24].

**Table 3 Characteristics of participants before and after the intervention regarding dependent variables (n=20)**

| Variables | Mean (SD) | | Min – Max | | Skewness | | Kurtosis | |
|---|---|---|---|---|---|---|---|---|
| | Pre-test | Post-test | Pre-test | Post-test | Pre-test | Post-test | Pre-test | Post-test |
| *Flanker task* | | | | | | | | |
| Accuracy scores | 116.45 (42.02) | 146.75 (13.49) | 26 – 160 | 115 – 160 | -0.93 | -1.28 | -0.50 | 0.76 |
| Congruent mean | 30.5 (10.53) | 37.45 (3.19) | 8-40 | 29-40 | -1.04 | -1.52 | -.42 | 1.75 |
| Incongruent mean | 26.25 (11.58) | 33.80 (6.61) | 0-40 | 17-40 | -.73 | -1.39 | -.35 | 1.14 |
| RT in ms | 605.67 (79.33) | 558.55 (55.09) | 487.08-735.07 | 470.13 – 644.66 | 0.16 | -0.38 | -1.78 | -1.31 |
| Congruent mean (ms) | 594.04 (82.62) | 553.62 (55.54) | 474.25-733.53 | 456.28-662.18 | 0.31 | -.15 | -.96 | -.71 |
| Incongruent mean (ms) | 645.88 (80.30) | 604.34 (74.04) | 517.90-794.40 | 487.55-705.40 | 0.95 | -.13 | -1.03 | -1.30 |
| *Digit span test* | | | | | | | | |
| Accuracy scores | 4.70 (2.08) | 6.85 (2.06) | 1 – 10 | 4 – 13 | 0.64 | 1.75 | 0.91 | 3.84 |
| RT in ms | 7214.53 (2078.73) | 9812.71 (10791.41) | 4061.10 – 12493.67 | 2472.00 – 54335.00 | 0.65 | 4.07 | 0.99 | 17.42 |
| RT in each memory span | 2047.09 (1578.95) | 1569.61 (1836.71) | 469.07-7265.00 | 395.33-9055.83 | 2.16 | 3.91 | 5.74 | 16.44 |
| *Berg Card Sorting test* | | | | | | | | |
| Percentage of correct responses | 61.05 (16.96) | 81.10 (5.96) | 30.47-88.79 | 64.84-88.68 | -.18 | -1.30 | -1.15 | 1.63 |
| Perseverative errors | 32.80 (24.44) | 15.15 (6.24) | 0.00-88.00 | 7.00-35.00 | .97 | 1.73 | .31 | 4.48 |
| RT in ms | 5143.48 (3261.92) | 3252.38 (1045.57) | 2326.60 – 12634.70 | 1907.95 – 5795.33 | 1.53 | 1.09 | 1.08 | 0.56 |

RT=response time    ms= millisecond

The results of the preliminary paired t-test analysis for inhibition, working memory, and mental flexibility of MCI participants after undergoing MU-COPART implementation are presented in Table 4. The findings indicate a significant increase in the average response accuracy of inhibition (t= -3.37, p < 0.001), along with a significant decrease in the average response time of inhibition (t=2.67, p < 0.05). Moreover, there was a significant improvement in the average response accuracy scores of working memory following MU-COPART implementation, as evidenced by a t-score analysis of -5.67 (p < 0.001). However, no significant changes were observed in the average response time of working memory or in the average response time within each memory span, with t-score analyses of -1.09 and 0.94 respectively (p > 0.05. Notably, the average response time of mental flexibility was significantly shorter post-implementation compared to pre-implementation (t= -5.74, p < 0.001). Additionally, there was a significant change in perseverative errors, as indicated by a t-score analysis of 3.38 (p < 0.05). There results underscore the positive impact of MU-COPART implementation on cognitive function among MCI older participants (t = 2.85, p < 0.05).

**Table 4 Comparison of executive function in MCI participants before and after undergoing MU-COPART program (n=20).**

| Variables | Mean (SD) | Std.Error | t-test | df. | p-value |
|---|---|---|---|---|---|
| *Flanker task- Inhibition* | | | | | |
| Accuracy scores | -30 (39.85) | 8.91 | -3.367 | 19.00 | .003* |
| Congruent mean | -6.95 (10.24) | 2.29 | -3.035 | 19.00 | .007* |
| Incongruent mean | -7.55 (10.63) | 2.38 | -3.175 | 19.00 | .005* |
| RT in ms | 47.12 (79.00) | 17.66 | 2.668 | 19.00 | .015* |
| Congruent mean (ms) | 40.42 (85.54) | 19.13 | 2.113 | 19.00 | .048* |
| Incongruent mean (ms) | 41.54 (87.73) | 19.62 | 2.118 | 19.00 | .048* |
| *Digit span test- Updating* | | | | | |
| Accuracy scores | -2.15 (1.69) | 0.38 | -5.67 | 19.00 | 0.00** |
| RA in ms | -2598.18 (10706.88) | 2394.13 | -1.09 | 19.00 | 0.29 |
| RA in each memory span | 477.48 (2269.73) | 507.53 | 0.94 | 19.00 | 0.36 |
| *Berg Card Sorting test-shifting* | | | | | |
| Percentage of correct responses | -20.05 (15.63) | 3.50 | -5.74 | 19.00 | 0.00** |
| Perseverative errors | 17.65 (23.33) | 5.22 | 3.38 | 19.00 | 0.00** |
| RT in ms | 1891.10 (2962.62) | 662.46 | 2.85 | 19.00 | 0.01* |

**p < 0.001, *p < 0.05; RT=response time   ms= millisecond

*Discussion*

This impact study aimed to elucidate the effects of sustained participation in physical activity on EFs, specifically inhibition, working memory, and mental flexibility among the community-dwelling older adults with MCI. The findings revealed that regular engagement in physical activity using MU-COPART significantly improve the average response accuracy across tasks such as the Flanker task, digit span test and Berg card sorting test. These results underscore the effectiveness of continuous participation in physical activity through MU-COPART in maintaining and improving EF efficiency among community-dwelling older adults with MCI, particularly in terms of inhibitions, working memory, and mental flexibility.

Nonpharmacological interventions targeting cognitive decline in older adults emphasize exercise and the maintenance of physical activity as impactful strategies for preventing cognitive decline and delaying the clinical diagnosis of Alzheimer's dementia [14]. Previous studies have demonstrated that regular exercise has significant benefits on memory and EFs in older adults with MCI [25, 26], which is a prodromal stage of dementia [27], along with community-dwelling older adults with MCI, has also shown positive effects on cognitive performance related to adherence to exercise sessions [28-31].

For rehabilitation purposes, virtual reality (VR) or augmented reality (AR) technologies are promising tools for enhancing cognitive function and facilitating memory exercises for individual participants [32]. Several intensive studies have reported that physically, when combined with VR-based cognitive interventions, can activate specific brain areas and enhance general cognitive performance, EF, attention, and memory [33-35].

Moreover, VR and AR technologies have demonstrated potential in facilitating individual behavioral changes through the inhibitory learning approach embedded in VR and AR platforms [36]. Research indicates that VR-cognitive interventions have a significantly protective effect against MCI, aiding in maintaining functionality in older adults with MCI and potentially delaying the onset of Alzheimer's dementia [14]. VR enables the creation of realistic spatial and temporal scenarios resembling everyday life conditions. AR technology offers engaging, enjoyable, and multifaceted tasks for individuals, thereby influencing participant motivation [37]. The use of virtual-based interventions has proven to be an effective therapeutic strategy for enhancing EFs in older adults without cognitive impairment [38]. In particular, for older adults experiencing cognitive dysregulation, VR can provide a naturalistic environment, enhancing the biological validity of EF interventions. VR offers dynamic considerations for cognitive training that closely mimic real-world situations, tailored to the characteristics and needs of patients [37]. This enables individuals with MCI or dementia to learn or maintain their EFs [12, 39] within context while performing ADL. Obviously, EFs play a crucial role in performing ADLs. Therefore, VR or AR has been selected as an early intervention strategy to mitigate the effects of executive impairments and enhance everyday life functioning. In virtual environment (VEs), participants can dynamically interact with programmed activity, leading to complex emotional and cognitive experience that involve planning, organization, attention, and mental flexibility, mirroring real-life situations [37]. As a result, individuals with MCI can maintain a satisfactory quality of life alongside their cognitive abilities.

*Conclusion*

Overall, the findings from this preliminary pilot study show that continuous engagement in physical activity using MU-COPART significantly improves EFs, particularly inhibition, working memory, and mental flexibility of community-dwelling older adults with MCI in Thailand.

*Limitation and suggestion for further study*

Future research should aim to address limitations in this study. While the MU-COPART program has been designed as a home-based intervention for older adults, particularly those with MCI, the installation and utilization process involving a laptop computer and digital camera may pose challenges for older adults, especially those with cognitive impairment. Therefore, there is a need to develop the MU-COPART program into a smartphone application, which would enhance accessibility and usability for a wider range of individuals seeking health promotion opportunities in the future.

*Acknowledgements*

The authors would like to extend their gratitude to the community-dwelling older adults who participated in this study, whether through the health promotion hospitals or the senior citizens schools in Bangkok, Chonburi, and Nakhon Pathom, Thailand. They would also like to express their deepest appreciation for the facilities made available to us by the Faculty of Physical Therapy at Mahidol University and the College of Research Methodology and Cognitive Science at Burapha University.

*Conflict of interest*

The authors declare that they have no known conflicts of interest. The funder was not involved in the design of the study. Each author is individually responsible for the writing and content of this article.

*Funding*

The present study's funding was gained from the National Research Council of Thailand (Grant no. วช.อว.(อ)(กบท2)/213/2564). Their support is hereby gratefully acknowledged.